# Large barocaloric effect with high pressure-driving efficiency in hexagonal MnNi$_{0.77}$Fe$_{0.23}$Ge alloy

Qingqi Zeng[1,2], Jianlei Shen[1,2], Enke Liu[1,3*], Xuekui Xi[1], Wenhong Wang[1,3], Guangheng Wu[1], Xixiang Zhang[4]

1. Institute of Physics, Chinese Academy of Sciences, Beijing 100190, China
2. University of Chinese Academy of Sciences, Beijing 100049, China
3. Songshan Lake Materials Laboratory, Dongguan, Guangdong 523808, China
4. Physical Science and Engineering Division (PSE), King Abdullah University of Science and Technology (KAUST), Thuwal 23955-6900, Saudi Arabia.

**Abstract**

The hydrostatic pressure is expected to be an effective knob to tune the magnetostructural phase transitions of hexagonal MM'X alloy. In this study, magnetization measurements under hydrostatic pressure were performed on a MM'X martensitic MnNi$_{0.77}$Fe$_{0.23}$Ge alloy. The magnetostructural transition temperature can be efficiently tuned to lower temperatures by applying moderate pressures, with a giant shift rate of -151 K GPa$^{-1}$. A temperature span of 30 K is obtained under the pressure, within which a large magnetic entropy change of -23 J kg$^{-1}$ K$^{-1}$ in a field change of 5 T is induced by the mechanical energy gain due to the large volume change. Meanwhile, a decoupling of structural and magnetic transitions is observed at low temperatures when the martensitic transition temperature is lower than the Curie temperature. These results show a multi-parameter tunable caloric effect that benefits the solid-state cooling.

In recent years, there have been much interests in hexagonal MM'X (M and M' are transition metals and X is main group element) alloys due to the realization of magnetostructural transition[1], magnetocaloric effects (MCE)[2,3], negative thermal expansion[4] and magneto-strain[5] near room temperature in these systems. The hexagonal MM'X alloy MnNi$_{0.77}$Fe$_{0.23}$Ge was reported as a system with martensitic and magnetic transitions coupled together, which occurs at about 266 K on cooling from paramagnetic (PM) parent phase with Ni$_2$In-type structure to ferromagnetic (FM) martensite phase with TiNiSi-type structure[1]. With the coupling of magnetic and structural degrees of freedom, a large MCE was reported on the transition. As we know, the MCE is intensively studied due to its applications in magnetic refrigeration, thermomagnetic motors, controlling of drug delivery and etc.[6-8]. It is desirable to find external parameters to tune or enhance the MCE. For hexagonal MM'X family, there is a

---

remarkable character, i.e., giant volume expansion with values of 2 ~ 5% during the structural transition upon cooling[1,3,9]. A hydrostatic pressure can drive this transition to lower temperatures as the mechanical energy gain ($\Delta P \cdot \Delta V$) in the total Gibbs freedom energy tends to shift the equilibrium point of the phase transition. Thus, besides the chemical substitution[10,11], vacancy[12,13], magnetic field[14] and strain[15], the hydrostatic pressure is widely used[10,11,16-21] to tune a phase transition. In fact, as a clean means the pressure can be used to induce many physical behaviors[22-26]. The studies on MCE include not only a wide temperature range, but also a shift in transition temperature by external parameters, such as hydrostatic pressure, which can tune the transition to the required working temperature[21].

In this study, we investigate the effect of hydrostatic pressure on magnetism, martensitic transition and MCE based on a hexagonal MM'X alloy $MnNi_{0.77}Fe_{0.23}Ge$. The magnetostructural transition temperature can be efficiently tuned to lower temperatures, with a driving efficiency of -151 K GPa$^{-1}$. A temperature span of 30 K is obtained under the pressure, within which a magnetic entropy change of about -23 J kg$^{-1}$ K$^{-1}$ for a magnetic field change of 5 T is observed. A decoupling of structural and magnetic transitions occurs when the martensitic transition temperature is lower than the Curie temperature of parent phase at higher pressures. These results indicate that the caloric effect of this system can be tuned by multiple external parameters.

Polycrystalline ingots were prepared by arc melting high-purity metals under Argon atmosphere. The ingots were then annealed at 1123 K for 5 days and slowly cooled at a rate of 1 K min$^{-1}$ to room temperature to avoid stress in samples. The structure of measured sample was examined by x-ray diffraction (XRD) using Cu *Kα* radiation. Magnetization measurements were performed using a Quantum Design superconducting quantum interference device vibrating sample magnetometer (SQUID-VSM) equipped with pressure cell in fields up to 5 T in the temperature range of 5-300 K and for pressures up to 1.280 GPa. Daphne 7373 was used as the pressure transmitting medium. In order to determine the pressure in the pressure cell more accurately, a piece of Pb was loaded together with the proper size bulk sample as the superconducting transition temperature under certain pressure for Pb is known[27].

At ambient pressure, $MnNi_{0.77}Fe_{0.23}Ge$ undergoes a martensitic transition at about 266 K according to the previous report[1]. The structures of high temperature parent and low temperature martensite phase are shown in Figs. 1(a1) and 1(a2). The parent phase crystallizes in a $Ni_2In$-type hexagonal structure with space group of *P*6$_3$/*mmc* (194) while martensitic phase TiNiSi-type orthorhombic phase with *Pnma* (62)[28-30]. The volume change during the transition is about 2.7%[1]. In the right panel of Figs. 1(a1) and 1(a2) we sketch the atom position relation of two phases. The structural transition occurs via distortions of Ni/Fe-Ge hexagonal rings and Mn–Mn zigzag chains[3]. Magnetic field dependence of the magnetization (*M-H*) curves at 5 K under several pressures are shown in Fig. 1(b). A metamagnetic behavior occurs at about 0.4 T under low pressures (0 and 0.117 GPa), which is also consistent with the previous report[1]. The martensite phase in Mn-Ni-Fe-Ge system will form a spiral antiferromagnetic (AFM) structure under zero field at low temperature[1,2]. In addition, the *M-H* in this system shows metamagnetic transition from spiral AFM to FM at 5 K. In a different fashion, the *M-H* curves under 0.776 and 1.280 GPa show FM behavior of parent phase, indicating the suppression of martensitic transition by high pressure. Martensite phase exhibits higher saturation field and magnetization than those of parent phase. The former characteristic may be due to the higher anisotropy in the martensite phase. The different saturation magnetization is related to the different distances between neighboring Mn atoms in two phases[18,31,32]. However, the pressure-induced changes in magnetization in both phases (parent and martensite) are small, suggesting a minor variations of the exchange interaction, which may be attributed to a slight modification of the electronic density of states at the Fermi level under the present pressures[11].

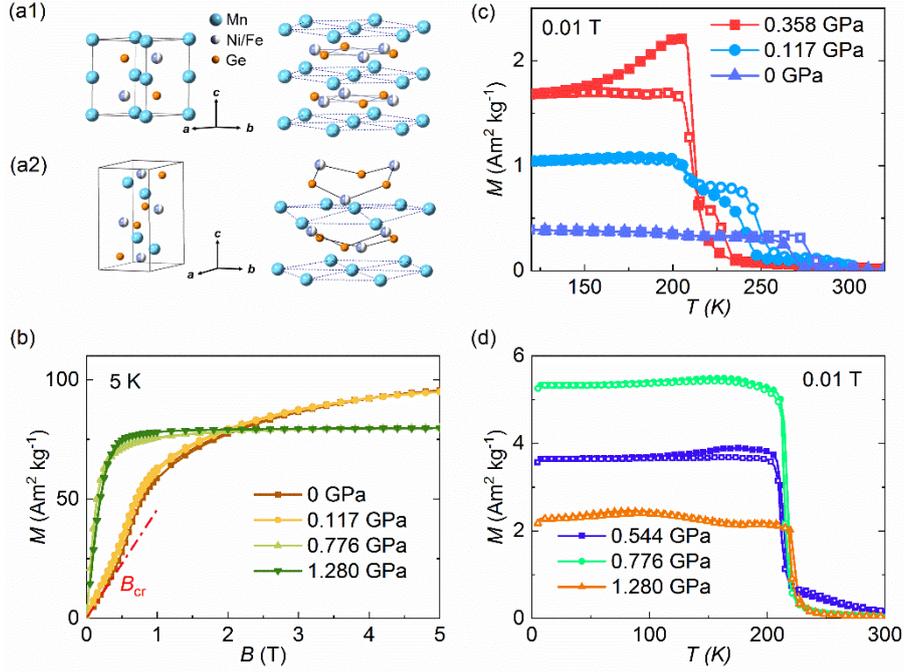

Fig. 1 (a1) and (a2) Primitive cell and a structure sketch of hexagonal parent and orthorhombic martensite phases. (b) Magnetic field dependence of the magnetization (*M-H*) curves at 5 K under different pressures. (c) and (d) Temperature dependence of the magnetization (*M-T*) in a magnetic field of 0.01 T under different pressures. The solid and open symbols represent the cooling and heating process, respectively.

The variation of magnetization and martensitic transition behaviors under different pressures are depicted by temperature dependence of the magnetization (*M-T*) in a magnetic field of 0.01 T, as shown in Figs. 1(c) and 1(d). The *M-T* curves under 0 and 0.117 GPa show clear temperature hysteresis which is associated with the forward and reverse martensitic transition temperatures, indicating a first order nature of the transition. The martensitic transition temperature is determined by the extreme point of the first order differential of *M-T* curve. The forward martensitic transition temperature ($T_M$) at ambient pressure is 267 K, which is consistent with previous report[1]. At $T_M$, while it is a magnetostructural transition, the parent phase in PM state transforms to the martensite phase in FM order. There is a magnetization change without hysteresis around 209 K under 0.117 GPa, which should be the Curie temperature of residual parent phase. The change in the magnetization between parent and martensite phases becomes larger with increasing pressure. Such an increase is a consequence of the lower martensitic transition temperature, at which the magnetization of FM martensite increases.

The *M-T* curve under 0.358 GPa is a little complicated due to the near decoupling of structural and magnetic phase transitions. When the temperature decreases, one can see the Curie temperature of parent phase ($T_C^A$) and entangled martensitic transition temperature at about 212 K. However, due to the lager span of structural phase transition than magnetic one, we can still see a "tail" of the martensitic transition. Thus there is a decrease of magnetization soon after $T_C^A$ and $T_M$ due to the lower magnetization of martensite compared to parent phase in a low magnetic field of 0.01 T. On the heating process, the first drop of magnetization is associated with the Curie temperature of residual parent phase, while the second drop at about 228 K is due to the reverse martensitic transition from martensite to PM parent phase. The *M-T* curves show only a transition without hysteresis above 0.544 GPa (Fig. 1(d)), indicating the occurrence of the suppression of martensitic transition by higher pressures, which is consistent with the above analysis about the *M-H* curves. The martensitic transition is shifted to lower temperatures by pressure and finally vanishes at about 210 K. As this temperature is relatively high, there should be other factors that suppress the structural transition besides the high pressure.

The variation of magnetic order of parent phase should be one reason. It seems that the martensitic transition will vanish when the FM transition of parent phase happens first during temperature decreasing. This suppression behavior has been observed in other MM'X compounds[1,9,12]. The interpretation could be that the parent phase is more stable at the FM state than the PM one. Thus the energy barrier between FM parent and martensite phase becomes difficult to overcome. In the present case, this suppression behavior occurs under the condition of the hydrostatic pressure loading. Note that the sudden drops of magnetization at very low temperatures in Fig. 1(d) are associated with the superconducting transitions of Pb as a pressure probe.

Figure 2(a) shows the pressure dependence of martensitic transition temperature and Curie temperature of parent phase. $T_M$ and $T_C^A$ are determined by the extreme points of the first order differential of the *M-T* curve. The large error bars under 0.321 and 0.330 GPa are due to the entanglement of the structural and magnetic transitions. As one can see, $T_M$ decreases while $T_C^A$ increases with increasing pressure. An increasing pressure leads to a shrinking of the cell volume and enhanced exchange interactions between moments on Mn atoms, which result in the rising of $T_C^A$. The pressure dependence of Curie temperature shows a similar behavior to MnCoGe system[33]. The application of hydrostatic pressure stabilizes the hexagonal parent phase because there is a big expansion of the volume during the martensitic structural transition upon cooling, which is common in hexagonal MM'X alloys.[3] Pressures above 0.544 GPa eventually suppress the martensitic transition.

The shift with pressure of $T_M$ is approximately calculated to be -151 K GPa$^{-1}$ in MnNi$_{0.77}$Fe$_{0.23}$Ge system, where $dT_t/dP = [T_M(0.358 \text{ GPa}) - T_M(0 \text{ GPa})]/0.358$ GPa. Figure 2(b) compares the shift of $T_M$ or $T_C$ (both are denoted as $T_t$) and the transition related magnetic entropy change ($|\triangle S_m|_{max}$) for a field change of 2 T in some systems. Here we do not distinguish the structural and magnetic transitions as we here study the capacity of pressure to tune the transitions, which is associated with the tuning of related magnetocaloric effects. Among all referenced systems here, MnNi$_{0.77}$Fe$_{0.23}$Ge exhibits prominent response of transition temperature with pressure. Therefore, for this alloy, it is easier to drive the martensitic transition by applying moderate hydrostatic pressures. Moreover, the large value of $dT_t/dP$ indicates that the magnetostructural transition could be induced reversibly between the $T_M$ at ambient pressure and the reverse martensitic transition temperature ($T_A$) at a proper pressure, for example, 0.117 GPa. Due to the above characteristic, MnNi$_{0.77}$Fe$_{0.23}$Ge as well as the related systems are proper platforms for researches of barocaloric effect and applications [36-38].

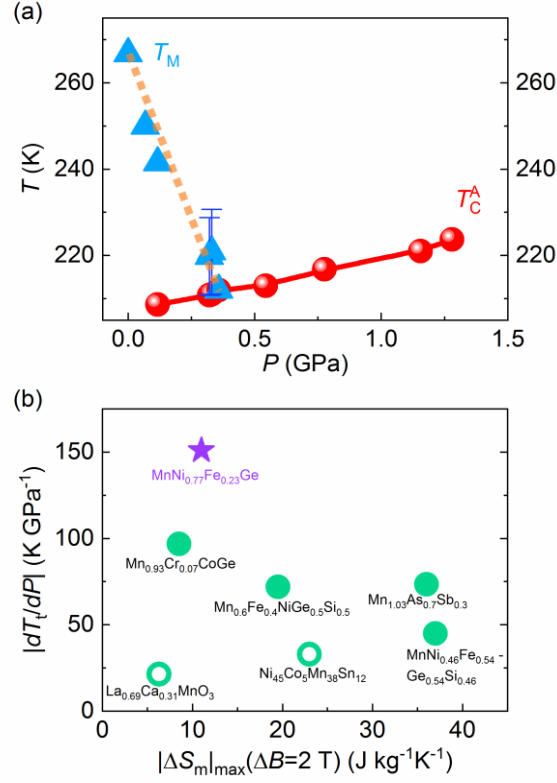

Fig. 2 (a) Pressure dependence of the martensitic transition temperature and Curie temperature of parent phase. The orange dot line is a guide of eyes. (b) The value of $|dT_t/dP|$ and $|\Delta S_m|_{max}$ (for a magnetic field change of 2 T) in several systems, where the temperatures of structural and magnetic transitions are denoted as $T_t$. The magnetic entropy changes are taken as the maximum values under different pressures for different systems. The solid symbols represent negative $dT_t/dP$ and $\Delta S_m$ while the open symbols positive values. Data are taken from references[11,17,19,21,34,35].

We performed magnetic isothermal measurements during cooling in a loop process[39] to avoid the influence of the magnetization history. The magnetic isothermals with temperature interval of -2 K are shown in Fig. 3.

The study of magnetocaloric effect (MCE) in MnNi$_{0.77}$Fe$_{0.23}$Ge system at ambient and under hydrostatic pressures was based on the Maxwell relation[39]. The magnetic entropy change associated with the magnetic field variation is in the following form,

$$\Delta S_m(T,B) = \int_0^B \left(\frac{\partial S}{\partial B}\right)_T dB = \int_0^B \left(\frac{\partial M}{\partial T}\right)_B dB \cdot \quad (1)$$

As the magnetic field induced martensitic transformation in MnNi$_{0.77}$Fe$_{0.23}$Ge is not reversible, i.e. the martensite phase induced by magnetic field loading will not transform to parent phase totally during demagnetization process. Thus we concentrate on the process which the parent phase transforms to martensite phase driven by increasing magnetic field. We calculated the magnetic entropy change of the magnetization process through Eq. (1) using magnetic isothermals data shown in Fig. 3. Figure 4(a) shows the entropy changes in several magnetic field variations under different pressures. The entropy change is reduced under 0.333 GPa and approaches nearly zero under 0.358 GPa due to the suppression of the martensitic transition by high pressures. However, the $\Delta S_m$ reveals a nearly unchanged value in the range of about −23.2 to −26.0 J kg$^{-1}$ K$^{-1}$ for a field change of 5 T (Fig. 4(b)) over the pressure range from 0 to 0.117 GPa. The conserved and tunable MCE under hydrostatic pressure can offer indications to the practical applications of caloric materials.

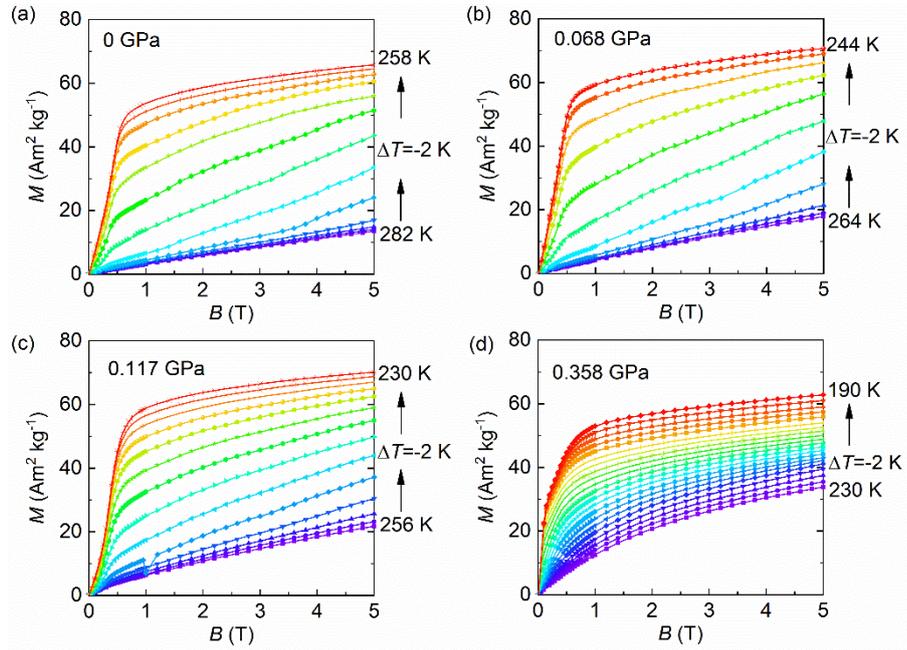

Fig. 3 Magnetic isothermals in a temperature interval of -2 K measured under 0, 0.068, 0.117 and 0.358 GPa.

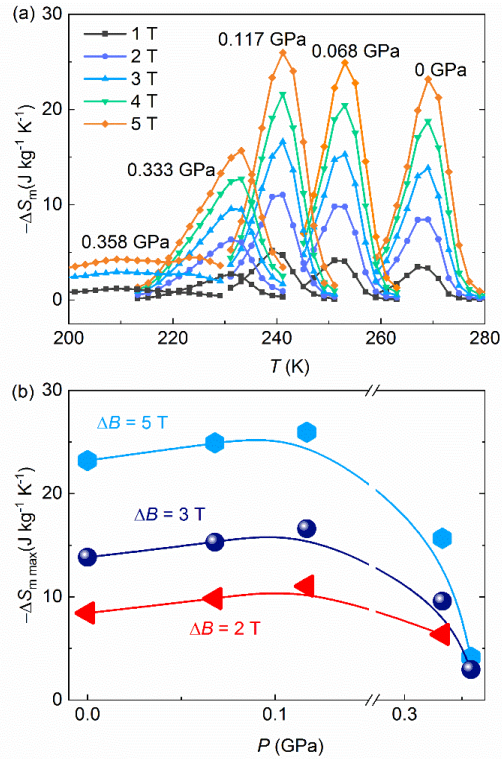

Fig. 4 (a) Temperature dependence of the magnetic entropy changes (-$\Delta S_m$) with magnetic field variation from 1 to 5 T under different pressures. (b) The maximum of the magnetic entropy changes (-$\Delta S_{m\ max}$) under different pressures for a field change of 2 T, 3 T and 5 T.

In summary, the study of magnetostructural transition and MCE under hydrostatic pressure in MnNi$_{0.77}$Fe$_{0.23}$Ge reveals that hydrostatic pressure can efficiently shift the magnetostructural transition temperature and thus the related MCE in this system. The decoupling of structural and magnetic phase transitions occurs at about 0.358 GPa, below which the MCE can be conserved and tuned by moderate pressure in a temperature range from 240 to 270 K. This characteristic can be useful in practical applications as one can tune the MCE to required temperature range by applying moderate pressures. More importantly, the remarkable ability to shift the magnetostructural transition temperature by hydrostatic pressure indicates a promising barocaloric effect driven by pressure loading around the magnetostructural phase transition in this alloy with a large volume change.

**Acknowledgements**

This work was supported by National Natural Science Foundation of China (No. 51722106), National Key R&D Program of China (No. 2019YFA0704904), Users with Excellence Program of Hefei Science Center CAS (No. 2019HSC-UE009), and Fujian Institute of Innovation, Chinese Academy of Sciences.

**References**

1. Liu E K *et al.* 2012 *Nat. Commun.* **3** 873.
2. Sun A *et al.* 2015 *Physica B* **474** 27.
3. Wei Z Y *et al.* 2015 *Adv. Electron. Mater.* **1** 1500076.
4. Zhao Y Y *et al.* 2015 *J. Am. Chem. Soc.* **137** 1746.
5. Xu K *et al.* 2017 *Sci. Rep.* **7** 41675.
6. Chen L *et al.* 2018 *Sci. China Phys. Mech. Astron.* **61** 056121.
7. Franco V *et al.* 2018 *Prog. Mater. Sci.* **93** 112.
8. Liu R S *et al.* 2020 *Chin. Phys. Lett.* **37** 017501.
9. Li Y *et al.* 2016 *APL Mater.* **4** 071101.
10. Nizioł S *et al.* 1983 *J. Magn. Magn. Mater.* **38** 205.
11. Caron L, Trung N T and Brück E 2011 *Phys. Rev. B* **84** 020414(R).
12. Liu E K *et al.* 2010 *EPL* **91** 17003.
13. Li Y *et al.* 2019 *Acta Mater.* **174** 289.
14. Shen B G *et al.* 2009 *Adv. Mater.* **21** 4545.
15. Moya X *et al.* 2013 *Nat. Mater.* **12** 52.
16. Fujita A *et al.* 2006 *Phys. Rev. B* **73** 104420.
17. Sun Y *et al.* 2006 *Appl. Phys. Lett.* **88** 102505.
18. Mañosa L s *et al.* 2008 *Appl. Phys. Lett.* **92** 012515.
19. Wada H, Matsuo S and Mitsuda A 2009 *Phys. Rev. B* **79** 092407.
20. Kaštil J *et al.* 2015 *J. Alloys Compd.* **650** 248.
21. Samanta T *et al.* 2015 *Phys. Rev. B* **91** 020401(R).
22. Khalid S, Sabino F P and Janotti A 2018 *Phys. Rev. B* **98** 220102.
23. Liu F *et al.* 2018 *Sci. China Phys. Mech. Astron.* **62** 48211.
24. You W *et al.* 2019 *Sci. China Phys. Mech. Astron.* **62** 957411.
25. Jiang S *et al.* 2019 *Chin. Phys. Lett.* **36** 046103.
26. Shang Y X *et al.* 2019 *Chin. Phys. Lett.* **36** 086201.
27. Eiling A and Schilling J S 1981 *J. Phys. F: Met. Phys.* **11** 623.

28  Johnson V 1975 *Inorg. Chem.* **14** 1117.
29  Bazela W *et al.* 1976 *Phys. Status Solidi A* **38** 721.
30  Fjellvåg H and Andresen A F 1985 *J. Magn. Magn. Mater.* **50** 291.
31  Bażela W *et al.* 1981 *Phys. Status Solidi A* **64** 367.
32  Liu E *et al.* 2011 *IEEE Trans. Magn.* **47** 4041.
33  Kanomata T *et al.* 1995 *J. Magn. Magn. Mater.* **140-144** 131.
34  Nayak A K *et al.* 2009 *J. Appl. Phys.* **106** 053901.
35  Taubel A *et al.* 2017 *J. Phys. D: Appl. Phys.* **50** 464005.
36  Lloveras P *et al.* 2015 *Nat. Commun.* **6** 8801.
37  Aznar A *et al.* 2019 *Adv. Mater.* **31** e1903577.
38  Lloveras P *et al.* 2019 *Nat. Commun.* **10** 1803.
39  Caron L *et al.* 2009 *J. Magn. Magn. Mater.* **321** 3559.